\documentclass[a4paper]{article}
\pdfoutput=1
\usepackage[utf8]{inputenc}
\usepackage{authblk}
\usepackage{tabularx}
\usepackage{graphicx}

\usepackage{booktabs}
\usepackage{adjustbox}
\usepackage{subcaption}
\usepackage{chemmacros}

\usepackage{hyperref}

\hypersetup{%
  pdftitle={Global property prediction: A benchmark study on open source, perovskite-like datasets},
  pdfauthor={Mayr, Felix \& Gagliardi, Alessio},
  pdfkeywords={DFT, perovskites, quantum machine learning, atomic fingerprints, property prediction},
  pdflang={en}
}

\usepackage[style=phys]{biblatex}
\bibliography{perovskite_paper.bib}

\title{Global property prediction: A benchmark study on open source, perovskite-like datasets}

\author[1]{Felix Mayr}
\author[1]{Alessio Gagliardi \thanks{alessio.gagliardi@tum.de}}
\affil[1]{Department of Electrical and Computer Engineering, Technische Universit{\"a}t M{\"u}nchen, M{\"u}nchen, Germany}

\newcommand{\software}[1]{\textsc{#1}}
\newcommand{\cmdl}[1]{\texttt{#1}}
\newcommand{\unit}[1]{\mathrm{#1}}

\begin{document}

\maketitle

\begin{abstract}
  Screening combinatorial space for novel materials - such as perovskite-like ones for photovoltaics - has resulted in a high amount of simulated high-troughput data and analysis thereof.
  This study proposes a comprehensive comparison of structural-fingerprint based machine-learning models on seven open-source databases of perovskite-like materials to predict bandgaps and energies.
  It shows that none of the given methods are able to capture arbitrary databases evenly, while underlining that commonly used metrics are highly database dependent in typical workflows.
  In addition the applicability of variance selection and autoencoders to significantly reduce fingerprint size indicates that models built with common fingerprints only rely on a submanifold of the available fingerprint space.
\end{abstract}

\section*{Introduction}
Perovskite-like materials are of paramount interest in the creation of novel photovoltaic devices.
While existing perovskite materials, such as \ch{CH3NH3PbI3} are unstable and/or contain toxic lead \cite{Niu2015, Boyd2018}, the available, combinatorial space of possible candidate compounds is extensive \cite{Saliba2019}.
This is especially interesting when considering mixtures and different structural phases, which might have widely varying properties \cite{Xiao2017,Konstantakou2017}.
Notably for binary mixtures of selected ions, it is already well established that the relation between am experimentally measured property (e.g. the bandgap) and material concentrations can be fit with simple, analytic functions \cite{Konstantakou2017, Gallardo2020}.
With the industry-led rise of machine-learning-(ML)-methods, there has been growing interest to predict such relationship in the high-dimensional space of all possible compounds using ML techniques \cite{Pilania2013, Zhang2020}.

While these approaches have been used for years in engineering and science in general \cite{Forrester2008}, the widespread application in computational materials science is relatively new and accompanied by the (re-)development of a wide range of ``fingerprinting functions'' \cite{Rogers2010,Behler2011,Rupp2012,Bartok2013, Schuett2014,Hansen2015,Faber2015,Huo2017,Faber2018,Stanley2019,Wang2020}.
These are necessary to encode the typical atomic and structural information describing materials of interest into a numerical vector format necessary for common ML-techniques.
For modeling computationally heavy quantum-chemistry calculations, two major approaches can be discriminated.
In the first, one tries to replace certain parts of already established frameworks with ML-models, e.g. the parametrization of molecular forcefields \cite{Behler2007,Smith2017} or the density functional in DFT \cite{Snyder2012}.
The second approach tries to create a surrogate model for prediction of materials properties given only the fingerprints as an input; typical properties for prediction with such a surrogate model are: stability/formation energy-terms \cite{Rupp2012,Faber2017,Seko2017,Stanley2019,Ye2018,Kaneko2019}, bandgaps \cite{Pilania2013,Pilania2016,Huang2018,Stanley2019,Wu2019,Saidi2020,Marchenko2020,Park2020} or even specific medication properties \cite{Stokes2020}.
Recent efforts also focus on the prospects of creating ``new'' materials from generative models or directly feeding the structural graph to a neural-network approximator \cite{Blaschke2017,Gomez-Bombarelli2018,Xie2018}.

This study focuses on the surrogate model approach applied to crystalline, perovskite-like materials.
In this field, most new methods or supposed performance improvements are only demonstrated with proprietary or novel datasets, severely limiting comparability to preexisting approaches and effectively hindering objective assessment of method performance across the field \cite{Marchenko2020,Im2019,Pilania2016}.
This is a direct result of the lack - to the author's knowledge - of a generally accepted, consistently annotated and high-quality benchmark database for crystalline materials, which could be used for benchmarking of new methods, like GDB-17 and its offspring QM9 are for organic systems \cite{Ruddigkeit2012, Ramakrishnan2014}\footnote{This might also be correlated to the prevalence of Kernel-Ridge-Regression (KRR) based methods \cite{De2016, Faber2017, Sutton2019}, which scale badly with very large (\textgreater 100k samples) databases, such as from the OQMD-project.}.
It should also be noted here that diverse databases - inevitably necessary for a complete surrogate model - tend to generate very large fingerprint vectors, which pose a theoretical and practical problem, when the size of the fingerprint is larger than the number of datapoints available for model building, possibly deteriorating performance \cite{Jain1982}. 

Most studies seem to implicitly employ both the regularizing properties of ridge-regression, as well as the (arbitrary) ``metric'' induced by a kernel function and do not warrant further attention to this problem \cite{Huo2017,Marchenko2020,Stanley2019}.

\begin{figure}[h]
  \includegraphics[width=\textwidth]{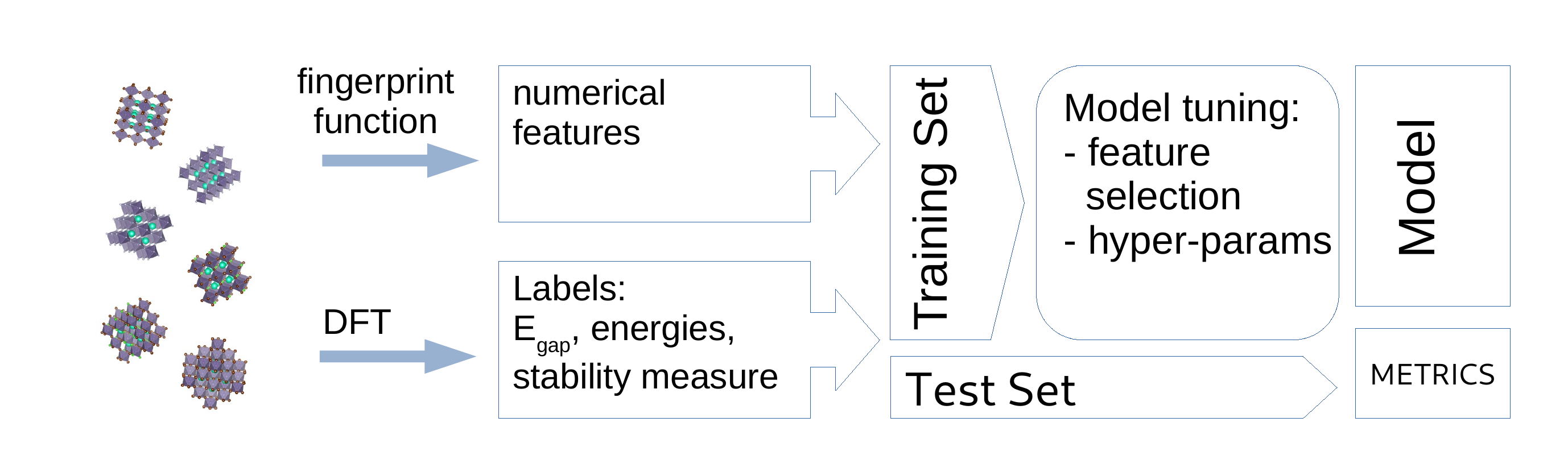}
  \caption{A typical materials prediction workflow for building a surrogate using structural fingerprints. The cost of DFT is typically magnitudes higher than for model evaluation or fingerprinting.}
  \label{fig:workflow}
\end{figure}

Herein, a typical materials science surrogate modeling approach (compare Figure \ref{fig:workflow}) employing Kernel-Ridge-Regression (KRR) is used on a variety of preexisting high-throughput databases of various crystalline, perovskite-like materials \cite{Castelli2012_cubic, Castelli2013_lowsymm, Pandey2018, Kim2017, Marchenko2020, Marchenko2020, Stanley2019}.
A host of different fingerprinting functions are compared \cite{Faber2015,Bartok2013,Huo2017}, including an improved, competitive version of the Property-Density-Distribution Function (PDDF)\cite{Stanley2019}.  

To assess the influence of KRR in squashing the dimensionality of the problem, this study employs a statistical feature selection process using variance thresholding and dimensionality reduction with neural-network autoencoders \cite{Stanley2019, Hinton2006}.

Rhe results underline that actual model accuracy as commonly published depends strongly on the dataset.
Intra-dataset even varying methodologies does yield comparable results within the estimated errors for bandgap predictions, while no single method is able to create equally accurate models for all datasets.
Analysis of the fingerprints reveals that models only rely on a subset of the available information in each at the given dataset scales.

\section*{Methods}
A typical property-predicting ML surrogate model for materials science is created in a supervised-learning setting on a sufficiently large set of (atomic structure, property)-tuples.
The arguably most simple way to do this, is to take basic compositional information, such as the fractional part of constituents, and then either fit a classic (non)-parametric model or train an artifical neural network (ANN).
A natural next step is to use means and higher moments of the property distribution over all atoms in a given structure as a numerical input vector, which yields suprisingly good results although it is completely insensitive to any structural differences \cite{Ward2017}.
For a constrained space of the input structures, which follow a given chemical structure (such as the perovskite-like \ch{ABX3}-one), one can also do this in a fine-grained ``per-site'' way and include basic structural information \cite{Saidi2020, Park2020}.

Although this has also been used for structurally diverse perovskites, this study will focus on fingerprints, which allow incorporation of structural information independent of a predetermined system.
Notably this includes the sine-matrix \cite{Faber2015}, the Smooth Overlap of Atomic Positions fingerprint (SOAP) \cite{Bartok2013,De2016} , the many-body-tensor-representation (MBTR) \cite{Huo2017} and the property density distribution function (PDDF) \cite{Stanley2019}, leaving out approaches only commonly employed with molecules and various adapations of local atomic symmetry functions \cite{Rogers2010,Behler2011,Rupp2012,Hansen2015,Faber2018}.

Except for the Coulomb-matrix-derived sine matrix \cite{Rupp2012}, all employed descriptors are derived from a shared basis, where for a given atom $j$, the environment is described by the atomic density of its neighbors $i$ (see \cite{Bartok2013, De2016})\footnote{NOTE: even for this very simple fingerprint formalism, one could replace the $\delta$-function with a more continous, Gaussian-like one, add specific weights for specific neighbor atoms $i$ and add a cutoff-function $f_c$, which limits the range, where $\rho > 0$.}:
\begin{equation}
  \rho_j(\vec{r}) = \sum_i \delta(\lvert \vec{r} - \vec{r}_i \rvert)\,.
  \label{eq:loc_env}
\end{equation}
Also, as this formalism is ``atom-centered'', any derived, numerical fingerprint is atom-local first and it is necessary to transform it to a ``global'' fingerprint to be used for predicting system-total properties for systems of varying compositions.
This transformation is done using special kernel-functions with kernel-based ML-techniques \cite{De2016} or by averaging the output over all atoms \cite{Himanen2020,Sutton2019,Stanley2019}.

While the SOAP fingerprint consists of the coefficients for expansion of the atomic density with radial and spherical basis functions, both MBTR and PDDF extend upon classic radial distribution functions.
Within the MBTR approach, both partial radial and angular distribution functions can be parametrized on different scales.
On the contrary, the PDDF weights contributions to a global RDF with atomic properties.
A thorough review of all used methods can be found in the SI.

A common problem with SOAP and partial RDF-based fingerprints is that they tend to generate large ($\mathcal{O}(1000)$) fingerprint vectors, which – when combined with a non-regularizing regression method – could lead to model overfitting if the dataset is also in the lower $\mathcal{O}(1000)$-range and require large amounts of computational ressources for kernel computation.
While sparsity of the individual fingerprints and the regularization part of KRR seem to alleviate this concern in a non-explicit way in most previous studies, this study employs variance-selection to methodologically shrink the input feature vector and observe the influence on surrogate modeling.

Additionally, dimensionality reduction techniques are employed to shrink the fingerprint vector, reducing the risk of overfitting and computational cost as well.
The underlying assumption  is that for most small-scale datasets, the structural and compositional variation within certain restrictions (such as: ``only pervoskite-like'' materials) is changing fingerprints in such a way, that this change can be projected onto a lower-dimensional manifold.

\begin{figure}[h]
  \includegraphics[width=\textwidth]{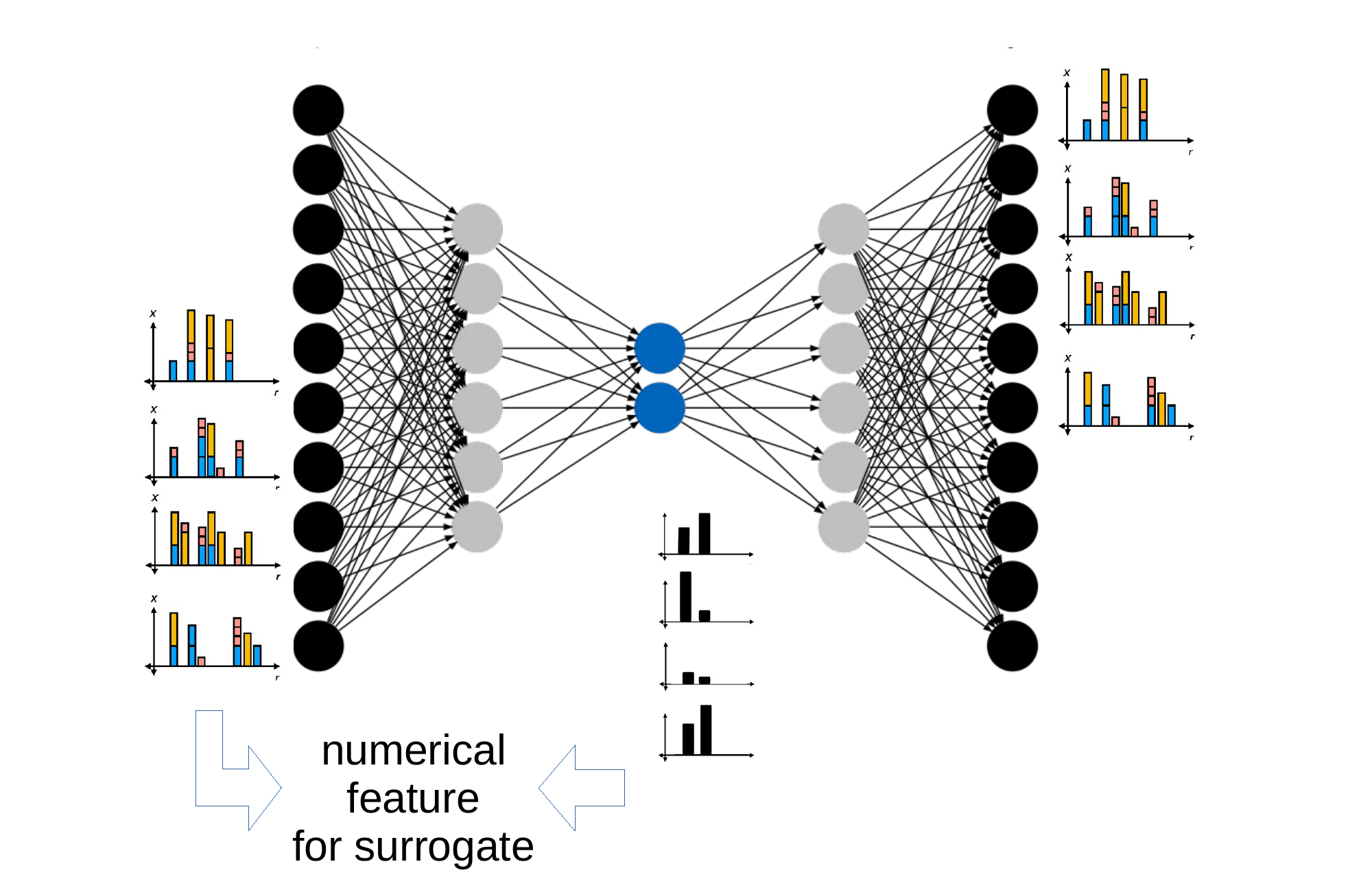}
  \caption{Network architecture of a 2 layer-autoencoder. The input is a numerical vector, the output tries to reconstruct the input, passing through a bottleneck, the so called ``latent space'' (blue). Surrogate models are then either built on top of the input or the intermediate representation.}
  \label{fig:AE}
\end{figure}

For this purpose this work proposes to use autoencoders, which are an unsupervised learning method using neural networks, passing the fingerprint as an input through an ``encoder'' network leading up to a ``latent''-layer (which is smaller than the original) and then up again trough a mirrored network (the ``decoder'') such that the original input is recreated (see Figure \ref{fig:AE}).
For building a regression model, the encoder then creates a compressed representation of all the fingerprints in question and these are fed to classical Kernel-Ridge-Regression.
Furthermore, one could - depending on the design of the study - feed all candidate compounds to the autoencoder including the ones, where one would like to do ML-based predictions (because training it is much cheaper than running full DFT for all).
Also, the reduced and ideally non-sparse feature-vector could allow to optimize for fingerprints (and subsequent structures) with a desired property within the restricted compositional and structural space of an numerical experiment \cite{Gomez-Bombarelli2018}.

\section*{Data}

The availability of consistent, high-quality data is crucial for building a ML model and eventual benchmarking of different modeling approaches.
Due to the lack of a shared, reproducible benchmark with a lot of common materials properties in the  ``solid-state''-community, authors tend to create their own datasets, when publishing new methods or researching new problems \cite{Marchenko2020,Im2019,Pilania2016}.
This process introduces the danger of the data being biased in an inadvertented way and thus giving unrealistic, non-generalizable results.
In addition, creating a suitable, high-quality DFT database of crystalline solids is a challenging task itself: one would like to have a high-fraction of ``physical'' systems, e.g. at their minimal energy, which requires extensive structural relaxations or even metadynamics to sample different likely substructures \cite{Kim2017}.
Calculations should also use a shared set of sufficiently exact parameters for all calculations to converge, which is hard to achieve with varying cell-sizes and some of the proposed inputs exhibiting metallic behavior.
Once a suitable amount of structures is relaxed, it still has to be assured that the model relates to physical reality, e.g. in the case of the bandgap as a property by incorporating spin-orbit-coupling and hybrid DFT, which generally seems to give bandgaps in good agreement with experiments compared to the underestimation by GGA \cite{Even2013,Hinuma2014a}.

While creating a high-quality database taking into account all these considerations is necessary to create a useful, physically exact surrogate model, for methodological development the usage of datasets of lower complexity is definitely possible.
Although a PBE-trained model might not yield accurate property values it can be expected that model accuracy will not get worse than the PBE baseline - which is still used for screening today – when trained on hybrid training data, while possibly leading to a significant performance increase.
Thus herein the choice fell on existing datasets of varying provenance and methodological backgrounds to assess whether the given methods are able to build effective surrogate models across different databases – a mutual theme is the inclusion of perovskite-like structures \cite{Castelli2012_cubic, Castelli2013_lowsymm, Pandey2018, Kim2017, Marchenko2020, Marchenko2020, Stanley2019}.

A large ($\approx 19\unit{k}$ samples) dataset of cubic perovskites is used from \cite{Castelli2012a, Castelli2012_cubic}.
It consists of cubic oxide-perovskite-scaffolds, featuring a wide range of cations and fractional replacement of the oxygen with flour, nitrogen and sulfur.
Optimized cubic-structures were found by scanning a range of lattice parameters and relaxing the resulting structure using DFT with the RPBE-functional.
For all non-metals, direct and indirect bandgaps were subsequently calculated with the GLLB-SC functional, which yields good agreement with experiments.
A subselection of these compounds (only with \ch{O} and \ch{N}-anion) and the same methodology were employed to derive a database of Ruddlesden-Popper layered-perovskites \cite{Castelli2013_lowsymm}.

Compared to these basic databases mainly varying the composition, the ``hybrid organic-inorganic perovskite-dataset'' by \cite{Kim2017} includes molecular cations \ch{A} in a ``classic'' \ch{ABX3}-halide-perovskite-scaffolds (with \ch{B}=\ch{Ge}/\ch{Sn}/\ch{Pb} and a halide \ch{X}).
Basic scaffold structures and cells were selected by running a minima-hopping simulation for initial \ch{ASnI3}-compounds resulting in a large number of different structural motifs, replacing the other sites and running a structural relaxation with the rPW86-functional. Bandgaps were then evaluated with the final structures at the position of both the direct and indirect gap in the relaxation calculation using hybrid DFT (HSE06).

In addition, a database of \ch{A2BCX4}-type materials was selected \cite{Pandey2018}, which are similar in size and scope to typical double-perovskites.
Structures are based on six different prototypes with the composition determined by empirical rules. Structures were optimized using PBEsol, with meta-GGA then used for energy calculations and GLLB-SC for accurate bandgaps.

Two smaller databases based on plain GGA and simple relaxation of base structures are also included:
first, a recently published dataset based on experimentally available 2D-perovskite compounds \cite{Marchenko2020}.
The structures therein generally resemble surfaces and thus exhibit widely varying cell-sizes\footnote{the reader should be aware that this database is apparently being updated. Thus, the shown statistics only show a snaphshot prior to publication of this paper (2020-07-28)}.
Second the database used by the authors in the introduction of the PDDF, consisting of relaxed, lead-free, inorganic mixed $2 \times 1 \times 2$-cubic-cell perovskites calculated at the GGA-level using an LCAO approach even for bandgaps is included \cite{Stanley2019}.

As all these databases incorporate a wide variety of species, fingerprints treating each pair of possible species separately (SOAP, MBTR) might be at a disadvantage and thus the crystalline dataset used in the Nomad-2018-Kaggle-competition consisting of a wide variety of  \ch{(Al_x Ga_y In_{1-x-y})_2 O_3}-compounds was included as a further reference \cite{Sutton2019}.

A basic overview of core properties of all used databases is found in Table \ref{table:dbs}, including summary statistics over all datapoints for structure size and the type of the bandgap/energy property. Note that these properties are not comparable inbetween different databases.

\begin{table}
  \begin{tabularx}{\textwidth}{l|X|X|X|X|X|X|}
    \toprule
     & Total compounds & unique species & size & max cell vector [$\unit{\r{A}}$]  & avg. cell vector [$\unit{\r{A}}$]  & bandgap [$\unit{eV}$] \\
    \midrule
    Kim \cite{Kim2017} & 1346 & 11 & 15.1 [9-21] & 7.4 [4.4-11.7] & 6.0 [4.3-7.6]  & 3.8 [1.52-6.63] (HSE06)  \\\midrule
    Pandey \cite{Pandey2018} & 1341 & 25 & 19.8 [16-32] & 11.9 [7.0-21.5] & 7.7 [6.4-10.5]  & 2.1 [0.01-4.28] (GLLB-SC)\\\midrule
    Stanley \cite{Stanley2019} & 344 & 9 & 20 [20-20] & 11.6 [10.9-13.9] & 9.2 [8.7-9.9]  &  1.4 [0.35-3.08]  (LCAO, PBE)\\\midrule
    Castelli \cite{Castelli2013_lowsymm} & 1984 & 47 & 20.9 [14-44] & 10.6 [7.3-24.0] & 6.6 [5.5-9.4]  & 3.5 [0-8.44]  (GLLB-SC)\\\midrule
    Castelli \cite{Castelli2012_cubic} & 18928 [735 nonzero gaps] & 56 & 5 [5-5] & 4.1 [3.3-5.7] & 4.1 [3.3-5.7]  & 0.1 [0-7.90] (GLLB-SC) \\\midrule
    Marchenko \cite{Marchenko2020} & 445 & 16 & 48.8 [4-452] & 27.7 [9.4-102.0] & 13.2 [6.4-22.9]  & 2.4 [1.65-3.53] (LCAO, n/a) \\\midrule
    Sutton \cite{Sutton2019} & 3000 & 4 [4-4] & 61.7 [10-80] & 15.1 [9.0-28.0] & 9.0 [4.8-10.8]  & 2.1 [0.0-5.84] (PBE)\\\midrule
    \bottomrule
  \end{tabularx}
  \caption{Overview of the used databases. Size (number of atoms), length of the maximal/geometrical average cell-vector ($\unit{\r{A}}$) and bandgap are all given in the format: ``mean [minimum - maximum]''. In parenthesis, for the bandgap, the choosen exchange functional is given.}
  \label{table:dbs}
\end{table}

\section*{ML Experiments}

In order to faciliate the comparison objective, the property prediction workflow is standardized across all databases and no dataset-tailored parameters or methods beyond the statistical model fitting/training procedure are used (see Figure \ref{fig:workflow}).
First, each randomly shuffled dataset is split into a 80\%-set for training and validation, while the remaining 20\% is set aside for testing.
Then the chosen fingerprinting function is applied to the structures, feeding the output either to an intermediate step reducing the fingerprint with an autoencoder or variance selection, or directly building the model using the fingerprints and a selected global property as a target.
For all models, 5-fold cross-validation was used to tune hyperparameters of a Kernel-Ridge-regression model using radial basis functions.
Finally, the resulting model is evaluated on the test set, resulting in an estimation of prediction accuracy in Table \ref{table:gaps} for direct bandgaps and for per-atom (formation) energies for each compound (in the SI).
Each model is evaluated using the mean-absolute-error (MAE) metric to estimate the error of the prediction and the R²-score (coefficient of determination) to classify the adherence to the ideal (prediction = ground truth)-relation, as the MAE alone depends strongly on the dataset.
The MAE-metric was deliberately chosen over the root mean-squared-error (RMSE) used in similar works \cite{Pilania2016,Langer2020}, because it disemphasizes outliers in predictions and is independent of the sample size \cite{Willmott2005}.
Also for a materials prediction workflow, where the end result will be validated with high-level calculations or experiments from a relatively large array of surrogate-qualified candidates, singular predictions which are off by a large amount are less relevant.
The results shown in Table \ref{table:gaps} are the average of 10 different train-test splits with the standard deviation used as an error estimate.
In face of the small datasets and non-standardized train-test-splits this method was chosen to avoid sampling a pathological, non-generalizable split \cite{Stanley2019, Langer2020}.

While nothing precludes the use of neural networks or other regression methods, Kernel-Ridge-regression was used throughout all experiments due to its low number of tunable parameters and its popularity within previous work \cite{Langer2020, Stanley2019, Marchenko2020, Sutton2019, Faber2018, Huo2017}.
The ``meta''-kernel approach was evaluated as well, specifically for the SOAP descriptor, but ultimately discarded, as it requires an enormous amount of computational time for kernel evaluation, while only marginally improving results \cite{De2016,Himanen2020}.

Although there is a magnitude of ``global'', macroscopic properties available \cite{Faber2017}, the employed databases only include bandgap and energy measures.
While the bandgap can be used ``as-is'' as a global property and is comparable except for intrinsic differences in the methods accuracy between databases, energy measures vary, with the availability ranging from bare total DFT-energies to formation energies within different, non-comparable frameworks.
Remedying this would require recalculating all compounds in a shared framework, which is beyond the scope of this study.
Thus, the focus lies upon the bandgap prediction models, with performance of prediction models for different kinds of formation energies and intensive ``per-atom''-DFT-energies shown in the SI.

To assess a baseline performance level for the more advanced methods this study includes the results of a dummy-regressor, returning the mean of the training dataset for all ``predictions'' on the test set.
Only on the hybrid perovskite database \cite{Kim2017}, some handpicked features (8 features: avg., site-specific properties for the ions \cite{Pilania2016, Ward2017} ) were considered and show a relatively good model ($R^2\approx 0.79$) with a MAE of $\approx 380\,\unit{meV}$ for the bandgap.
At this point it becomes apparent that the MAE alone gives no real indication for the quality of a surrogate model.
For example the dummy-regressor on the Marchenko database \cite{Marchenko2020} achieves a similar ``performance'' as the primitive predictor on the Kim-database \cite{Kim2017}, which already improves significantly on the dummy-prediction there, both with the MAE and the R²-score.
With the bandgap prediction, creation of a decent ($R^2\simeq 1$) model for the full dataset of cubic perovskites \cite{Castelli2012_cubic} was not possible and thus the subset of perovskites with non-zero bandgaps was selected for modeling \cite{Pilania2016}.

For the SOAP-fingerprint, the sparse, single-constituent fingerprints of a crystal were taken and averaged to create a global descriptor \cite{Sutton2019,Himanen2020}.
The parameters used in fingerprint creation \footnote{these could be thought and optimized as hyperparameters of the whole ``machine'' – though this hinders general applicability and requires expensive remodeling for new data} were picked from the existing literature, where widely varying numbers for the modeled cutoff radius and the number of radial and spherical basis functions are given without any reasoning (see the SI for a listing) \cite{Bartok2013, De2016, Sutton2019, Marchenko2020}. Assuming that large systems (such as in \cite{Castelli2013_lowsymm, Marchenko2020}) might benefit from modeling a larger cutoff radius around individual atoms, the parameters from \cite{Marchenko2020} were also included with a radius of $16 \unit{\r{A}}$. \footnote{Note that the dataset in \cite{Marchenko2020} is proprietary and based on a subset of the available database, so the results are not directly comparable.}

Kernel-calculation for the full fingerprints is very compute-intensive with fingerprint size in the 5-digit range depending on the number of species included in the data.
Thus individual features were min-max-scaled to the $[0,1]$-interval and variance selection with a 0.01 threshold was employed to significantly reduce the fingerprint size before scaling the data to unit-variance and feeding the data to a KRR model using radial basis functions.
Resulting models match or exceed performance of the usage of the full-fingerprint, where a simple linear kernel and no scaling where used as the radial basis function kernel required more computational time and didn't improve accuracy.

For the MBTR, only the $k2$-part was used, as this already results in a sizable fingerprint of size $s^2\cdot b$, where $s$ is the number of species in the database, while $b$ is the number of discrete bins, used to discretize the fingerprint on the given cutoff-radius of $16\,\unit{\r{A}}$ \footnote{This specific radius was chosen so it captures the environment of the maximum ``whole cell'' for most compounds (compare with the cell-vector geometrical average and maximum unit-cell vector lengths in \ref{table:dbs}). Also it could be discretized conveniently for numerical experiments as a multiple of 2.}.
$b=10$ was chosen and worked well with both the partial-rdf-equivalent representation as well as discretization over the inverse-radius and no scaling applied before feeding the data to the radial-basis KRR-model.
Using both, the full MBTR including the angular parts, as well as setting $b=100$ for the k2-version \cite{Huo2017, Langer2020} did not produce improved results consistently and significantly increased computational time.
Applying the same variance selection process used with SOAP did not provide improved results either.

Finally for the PDDF, different discretizations were explored for a radius of $16\,\unit{\r{A}}$ and a total of 8 properties.
With an amount of 160 features discretized with $0.8\,\unit{\r{A}}$-bins and a gaussian spreading of $1 \,\unit{\r{A}}$, the PDDF already works in building a bandgap model for all datasets – except the cubic perovskites – when scaled to standard variance.
A finer discretization with $0.1\,\unit{\r{A}}$-bins for the PDDF results in markedly improved results, while increasing the number of features 8-fold (1280).
Using a simple, 1-layer linear-activation autoencoder architecture trained on the $[0,1]$-scaled PDDF-representation of the training data alone, allows encoding the fingerprint into a 160-feature representation again.
Using this representation with KRR consistently reaches the performance of the full representation hinting that the PDDF-fingerprint indeed represents a low-dimensional manifold describing the data.
Further studies could be conducted to explore whether and how the latent space is actually a representation of this manifold and how it relates to basic input structural data.
In a similar vein, Schrier \cite{Schrier2020} explored the eigenspectrum of the Coulomb-Matrix fingerprint for molecular data and found, that even this already shrunken representation can be further reduced.

Detailed results can be found in Table \ref{table:gaps} for the bandgap and the SI for energy predictions and the remaining SOAP and MBTR-related experiments.

\section*{Evaluation}

For all of \cite{Pandey2018, Kim2017, Sutton2019, Stanley2019}, both the PDDF-approach and SOAP yield comparable prediction accuracy below $120\, \unit{meV}$ MAE with a slight lead for the SOAP fingerprint.
In case of the PDDF, both increasing the number of discretization steps and using the weigthing proposed by \cite{Hemmer2007} considerably improve results compared to the original rediscovered approach \cite{Stanley2019}.
Against that, the results of the SOAP method seem relatively independent of parametrization in spherical and radial basis functions.
Prediction is neither changed by decreasing the smearing of the atomic positions (``+fine''-attribute), but for \cite{Pandey2018,Stanley2019} increasing the radius expanded in the fingerprint results in a marked improval.
The sine-matrix approach is only significantly improving on the dummy-predictions in the case of constant system size \cite{Sutton2019} or with the hybrid perovskite dataset incorporating a high number of atoms for all systems \cite{Kim2017}.
Except for \cite{Sutton2019}, all MBTR-parametrizations lag behind, regardless of the specific setup.
While all best-performing prediction MAEs are of similar magnitude, it is notable that the baseline differs: for \cite{Pandey2018, Kim2017, Sutton2019} the error of educated guessing is $\approx 800 \,\unit{meV}$, whilst it is only $\approx 300 \,\unit{meV}$ for \cite{Stanley2019}.

Conversely, for \cite{Castelli2013_lowsymm} the MBTR-representation discretized on the inverse-radius-grid shows the best results, albeit model quality measured with the $R^2$-coefficient doesn't reach the best results of the previously discussed datasets and the best MAE is nearly doubled to $250 \,\unit {meV}$.
Both SOAP and the PDDF are performing worse for this datasets.

Similar, for the large-cell data from \cite{Marchenko2020}, the PDDF approach performs worst, independent of parametrization.
With errors of $130 \,\unit{meV}$ and $140\, \unit{meV}$ respectively, SOAP and MBTR are leading, though compared to the other databases with comparable MAEs, this is a considerably smaller improvement on random guessing!

Finally for the cubic perovskites \cite{Castelli2012_cubic}, no model reaches a satisfactory $R^2$ even with the dataset reduced to non-metallic compounds only.
MBTR leads the field with an MAE of $700\,\unit{meV}$ followed by SOAP and the PDDF in $50\, \unit{meV}$ increments.
Here, the proposed methods for building a surrogate model seem to fail, possibly a result of the discontinous nature of the input structures just being the results of simple combinatorics.
Thus, for the sparse SOAP and MBTR-fingerprints, most features just are incomparable with some parts being non-zero only in singular samples.

\begin{figure}[h]
  \includegraphics[width=\textwidth]{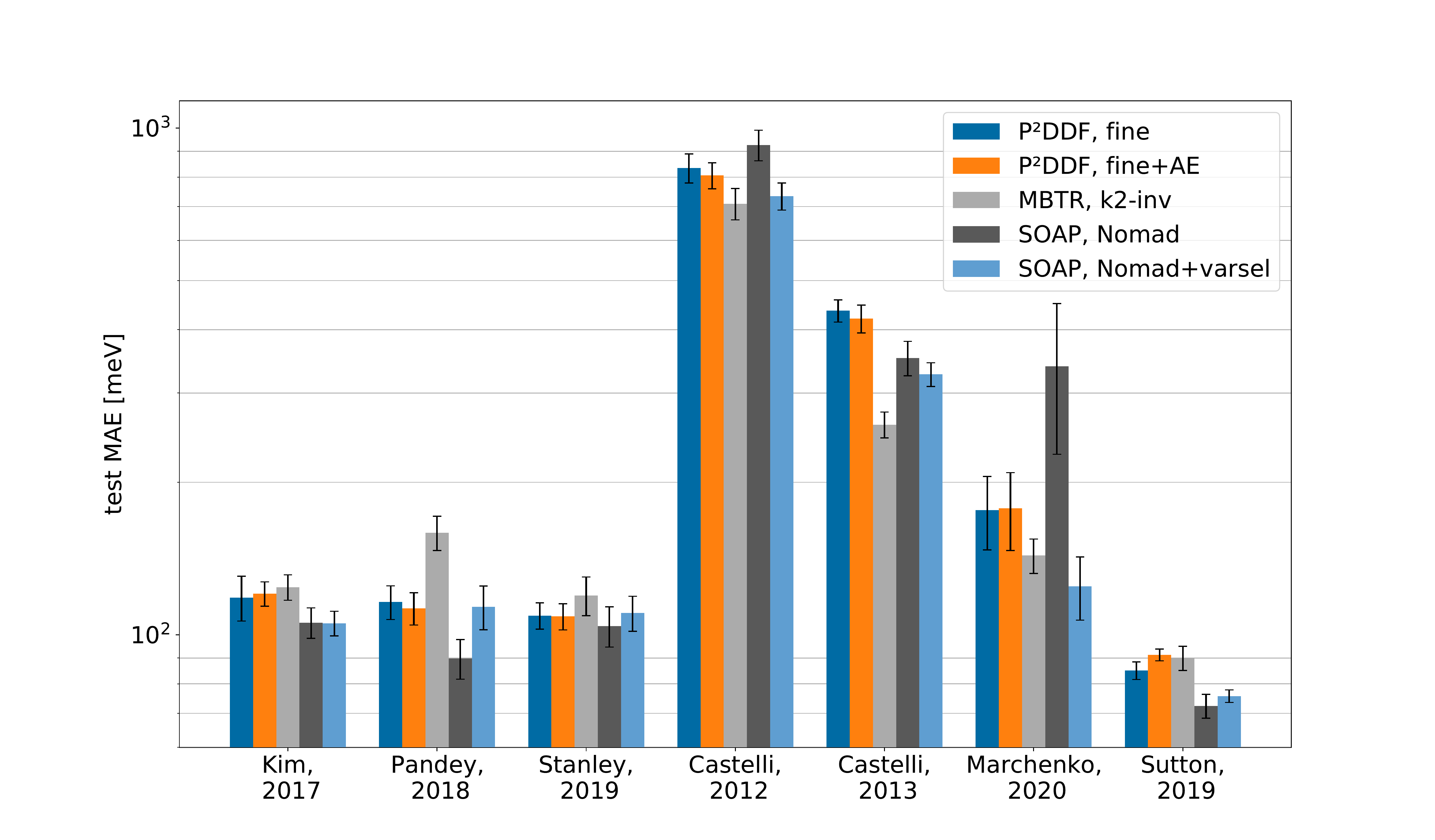}
  \caption{Bandgap prediction error visualized for selected model across all databases}
  \label{fig:gaps}
\end{figure}

Overall, as visualized in Figure \ref{fig:gaps} and further shown for all parametrizations in the SI, the exact choice of specific fingerprinting parameters or even the basic method has a much less produced effect on resulting errors than the choice of the database.
Even for technically very pathological parametrizations, e.g.: smoothing distributions with Gaussians of a similar width as the distribution range or the opposite for SOAP, the errors do not change on the order of magnitude.
While this study did not perform any large scale fingerprint hyperparameter tuning \cite{Langer2020} – instead choosing to replicate previous studies methodology, spanning a wide range of parameters – this indicates that for most practical screening applications the choice of method is less important than having a ``suitable'' database.
``Suitable'' in this case goes far beyond the addition of new datapoints, as the failure of building  a very good model for the data from \cite{Castelli2012_cubic} and \cite{Castelli2013_lowsymm} shows.
While one might attribute this to the high amount of unique species in these datasets (see Table \ref{table:dbs}), comparison between the similar results for the data from \cite{Kim2017}, \cite{Pandey2018}, \cite{Stanley2019} and \cite{Sutton2019} shows that this is not the only deciding factor.
This becomes especially apparent in the direct comparison of \cite{Kim2017} and \cite{Pandey2018}, where the number of available compounds is similar, yet the number of unique species is much higher in \cite{Pandey2018}.

\begin{table}
  
  \footnotesize
  \begin{center}
    \begin{adjustbox}{center}
    \begin{tabular}{lccccccc}
\toprule
{} & Kim \cite{Kim2017} & Pandey \cite{Pandey2018} & Stan. \cite{Stanley2019} & Cas. \cite{Castelli2012_cubic} & Cas. \cite{Castelli2013_lowsymm} & Mar. \cite{Marchenko2020} & Sutton \cite{Sutton2019} \\
\midrule
handpicked                 &             381±11 &                        - &                        - &                              - &                                - &                         - &                        - \\
dummy                      &             884±34 &                   730±19 &                   323±23 &                        1270±73 &                          1530±46 &                    332±15 &                   845±16 \\
sinematrix, eigenspectrum  &             368±15 &                   538±39 &                   212±15 &                        1088±77 &                          1102±60 &                    298±22 &                    141±8 \\
PDDF, basic                &             172±11 &                   199±13 &                   134±11 &                         930±80 &                           551±16 &                    179±11 &                    101±4 \\
PDDF, fine                 &              141±8 &                   139±14 &                   114±12 &                         888±57 &                           481±19 &                    176±20 &                     90±4 \\
PDDF, fine+AE              &              142±6 &                    143±7 &                   110±12 &                         879±61 &                           490±28 &                    170±19 &                     91±4 \\
P²DDF, basic               &             159±12 &                   172±14 &                   136±19 &                         888±69 &                           521±19 &                    207±32 &                     96±3 \\
P²DDF, fine                &             118±12 &                    116±9 &                    109±7 &                         834±55 &                           436±22 &                    176±29 &                     85±3 \\
P²DDF, fine+AE             &              120±7 &                    113±8 &                    109±6 &                         806±48 &                           421±27 &                    178±31 &                     91±2 \\
MBTR, k2-inv               &              124±7 &                   159±12 &                   120±11 &                         709±50 &                           260±15 &                    143±11 &                     90±5 \\
MBTR, k2-rdf               &              128±7 &                   144±13 &                   126±10 &                         786±57 &                           305±18 &                    140±18 &                     93±6 \\
SOAP, Marchenko            &              100±8 &                     85±9 &                    109±7 &                        1067±75 &                           349±27 &                    494±90 &                     70±5 \\
SOAP, De                   &              107±6 &                     97±9 &                    108±8 &                        1071±74 &                           329±25 &                    442±95 &                     78±4 \\
SOAP, Nomad                &              106±7 &                     90±8 &                   104±10 &                         926±64 &                           352±27 &                   339±112 &                     72±4 \\
SOAP, Marchenko, LR        &             110±11 &                     96±7 &                   122±10 &                       1288±130 &                           645±58 &                   939±330 &                     75±3 \\
SOAP, Marchenko+varsel     &              101±6 &                   111±10 &                    123±8 &                         738±52 &                           309±24 &                    132±17 &                     77±6 \\
SOAP, De+varsel            &              106±9 &                   112±10 &                    116±7 &                         777±50 &                           339±23 &                    135±20 &                     78±4 \\
SOAP, Nomad+varsel         &              105±6 &                   114±11 &                    110±9 &                         734±45 &                           327±18 &                    125±18 &                     76±2 \\
SOAP, Marchenko, LR+varsel &               99±8 &                     90±5 &                    104±8 &                         745±48 &                           324±27 &                    129±25 &                     76±3 \\
\bottomrule
\end{tabular}

    \end{adjustbox}
  \end{center}
  \caption{Results for predicting the calculated bandgaps for different methods, all results in $\unit{meV}$ for the mean-absolute error (MAE). The parameters for specific identifiers are listed in the supporting information. Note here, that P²DDF is used as a shorthand for the product-weighting proposed by \cite{Hemmer2007}. \cmdl{varsel} indicates that the machine learning was trained on the variance selected features of the specified fingerprint function}
  \label{table:gaps}
\end{table}

\begin{figure}[h]
  \includegraphics[width=\textwidth]{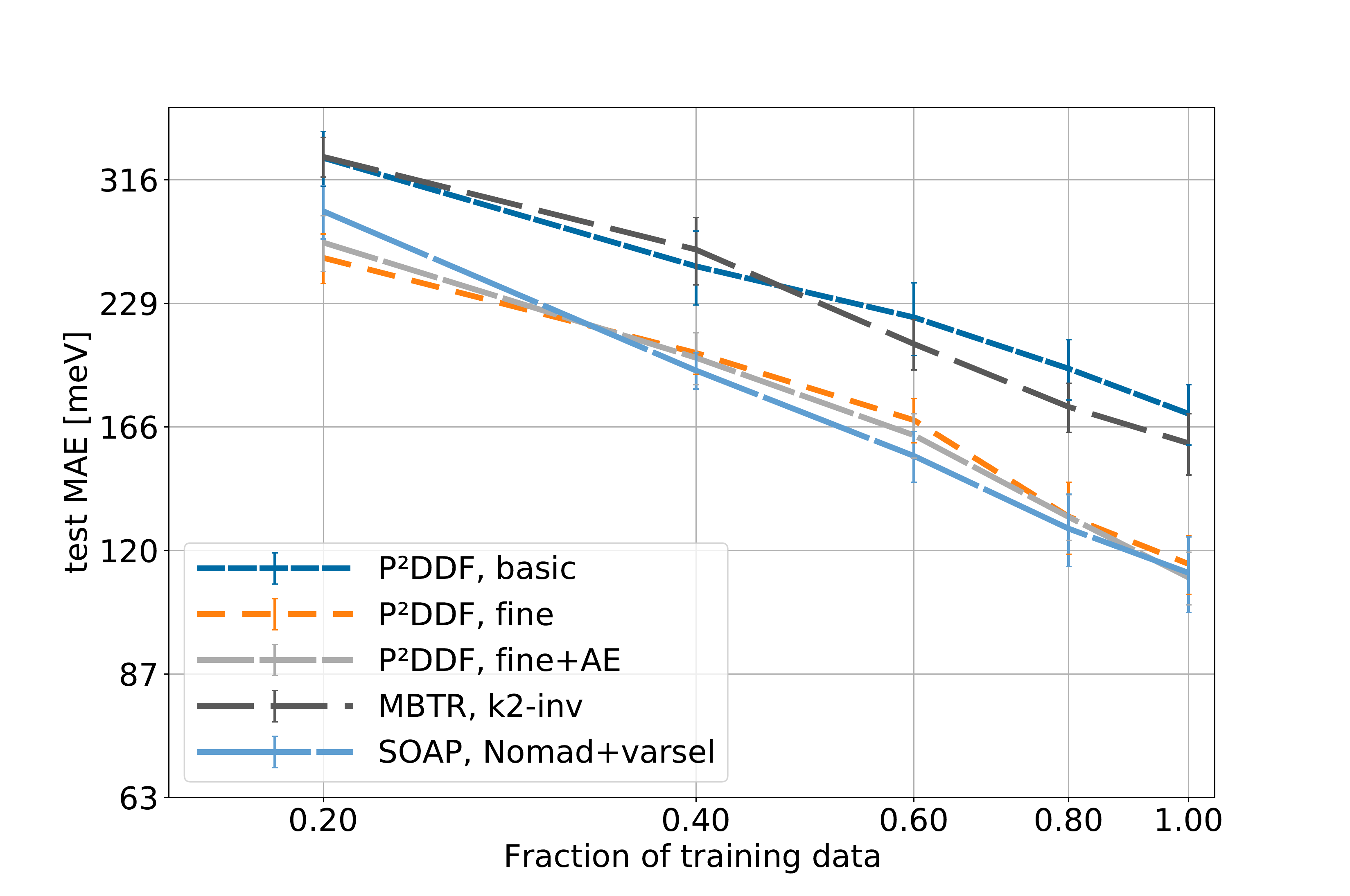}
  \caption{Learning curves for the data from \cite{Pandey2018}. Shown is the test MAE for selected machine learning models.}
  \label{fig:learning_curve}
\end{figure}

\begin{figure}[h]
  \begin{center}
  \includegraphics[width=0.5\textwidth]{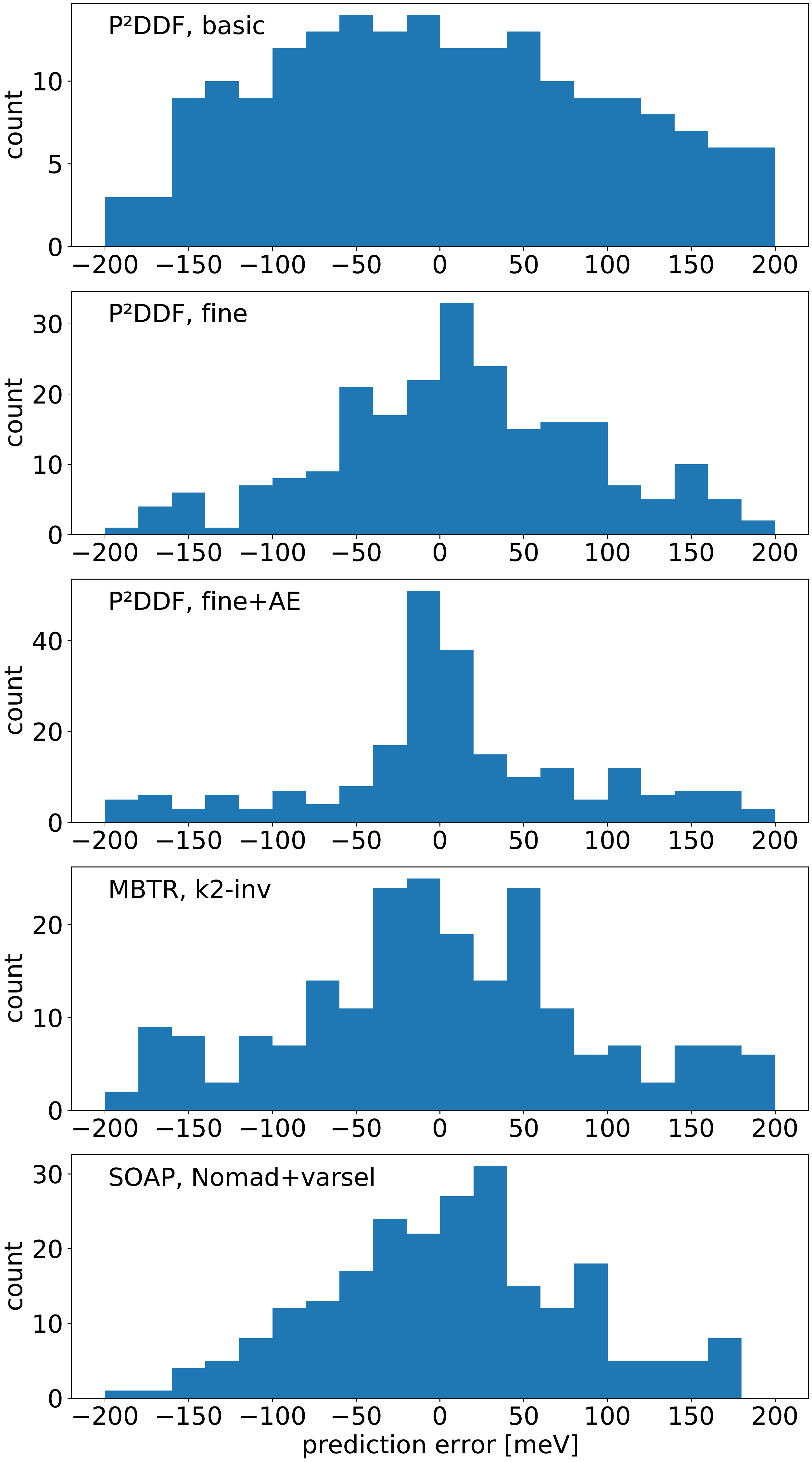}
  \caption{Error distribution for the best performing models for the data from \cite{Pandey2018} for selected  machine learning models}
  \label{fig:error_distribution}
\end{center}
\end{figure}

To provide further insight into model quality and limitations, Figure \ref{fig:learning_curve} and \ref{fig:error_distribution} provide learning curves and error-distributions for the bandgap prediction on the data from \cite{Pandey2018}.
The curves plot the average MAE of models evaluated on a $20\%$ test set versus the fraction of the respective training set used for creating the model.
All fingerprint methods and feature extraction techniques show a consistent improvement with increasing training data with no sign of flattening out, indicating that more training data could be used to further improve model quality.
In the low-data-regime, MAEs are exceeding $250 \,\unit{meV}$ and the models based on the autoencoded PDDF as well as the variance selected SOAP significantly trail the pure, small-bin PDDF.
By increasing the amount of data used for model creation, both SOAP and Autoencoder-model performance reach parity with usage of the full PDDF in model creation.
The autoencoder results could be related to a biased sampling which is not able to fully capture all structural features available in the dataset.
In a real prediction setting, one might thus include a larger array of prospective compounds, including the finally predicted datapoints for training the autoencoder.

Checking the error distributions for different best-case results with a 80/20-train/test-split shows gaussian like distributions, so there is no inherent bias of any of the tested modeling procedures (compare Figure \ref{fig:error_distribution}).

\begin{figure}[h]
  \centering
    \begin{subfigure}[t]{0.48\textwidth}
        \centering
        \includegraphics[width=\textwidth]{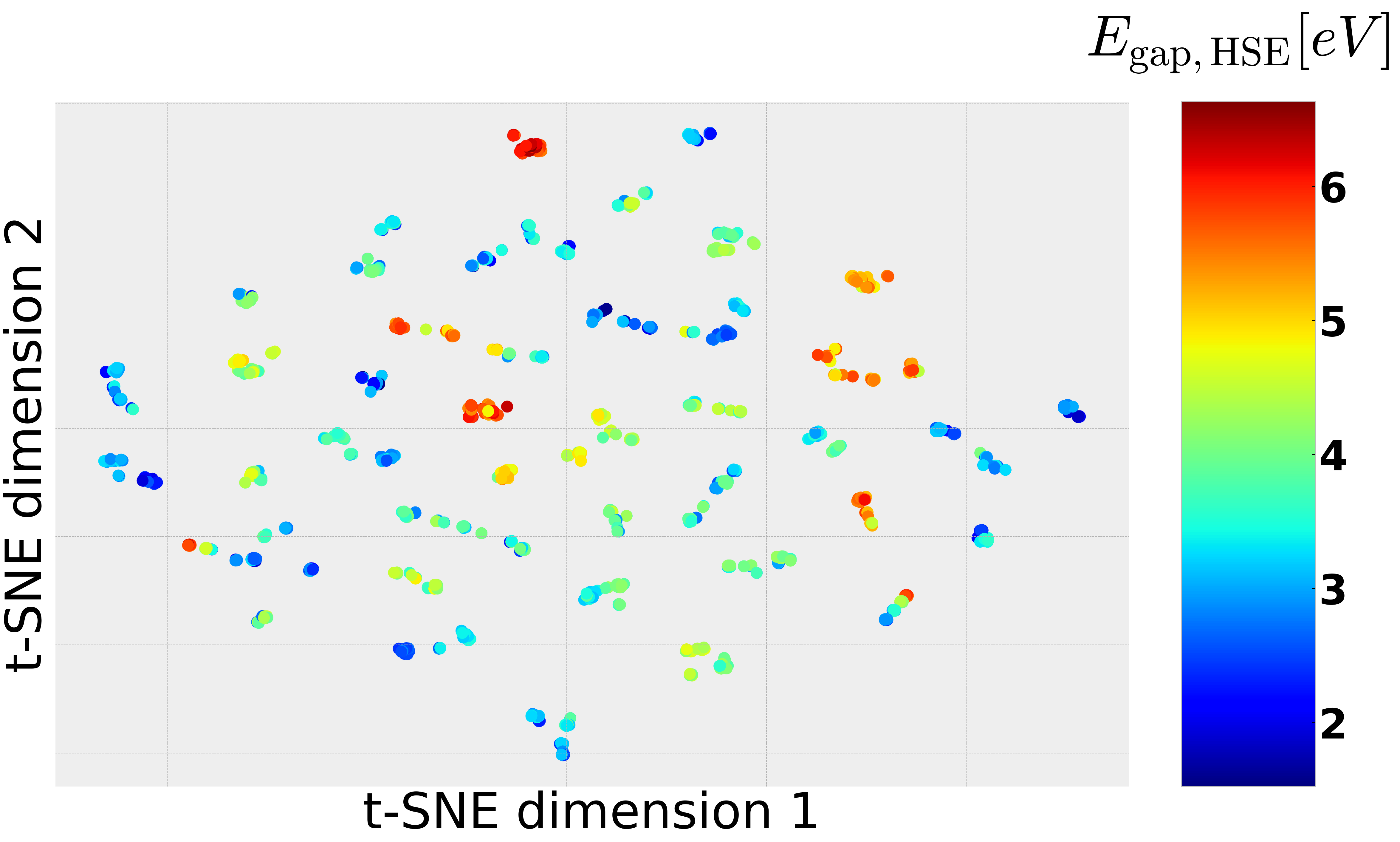}
        \caption{``raw'' fingerprint}
    \end{subfigure}%
    ~ 
    \begin{subfigure}[t]{0.48\textwidth}
        \centering
        \includegraphics[width=\textwidth]{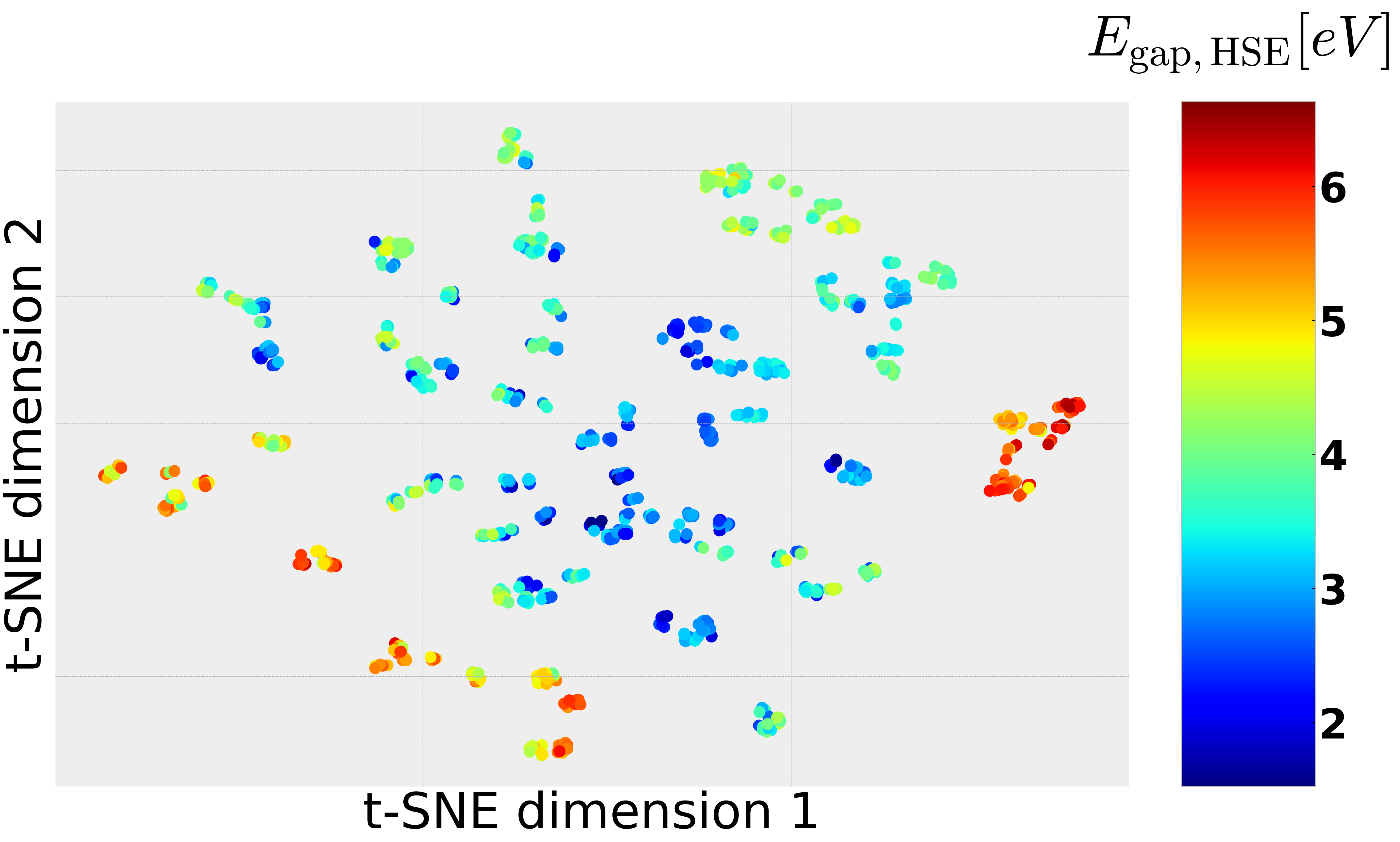}
        \caption{latent space of fingerprint-encoding autoencoder}
    \end{subfigure}
  \caption{t-SNE reduced PDDF fingerprint and its encoding in 2D. Bandgap overlayed and colorcoded to relate to physical realities.}
  \label{fig:explain_the_ae}
\end{figure}

Additionally, in a first effort to understand the effects of the autoencoder on the PDDF-fingerprint vectors used as input in the KRR model, t-SNE embeddings are used to create a 2-dimensional map of the relative ``neighborhoods'' accessible in the fingerprint (see Figure \ref{fig:explain_the_ae}) \cite{vanderMaaten2008,Wattenberg2016}.
The dataset from \cite{Kim2017} was used because it has a clear \ch{ABX3}-perovskite structure and a relatively well working model, so a relation to physical quantities is relatively easy.
When overlaying the bandgap on a plot of the first two t-SNE dimensions, it is evident that the autoencoder preserves information about the physical characteristics of the system and the resulting models are no statistical artifact compared to using the PDDF.
In the example, it even seems like the autoencoded representation is able to capture the bandgap-landscape in a much more continuous way than the original fingerprint, where a large number of singular high bandgap values are interspersed in the t-SNE-map.
This observation can be related to the fact that the autoencoded representation clusters depending on A and X-sites (see SI for the t-SNE-plots for A, B and X-site-''occupation''), with the B site not clearly distinguishable as separate clusters in 2D.
Conversely, the raw fingerprint does cluster mainly by the B and X-occupation, while the molecular ions at the A site are not distinguishable in 2D clusters.
As previous studies have shown that the B-ion is not very relevant for the bandgap \cite{Tao2019, Stanley2019}, this hints that the autoencoder might actually be able to extract a ``chemically informed'' representation from the fingeprints.
Obviously, the realizable advantage of this in building ML surrogates may be limited, as these are generally built on a space with a much higher dimension and the model can exploit more complicated relations than visualizable in a 2D map.

\section*{Conclusion}

The key takeaway of this study is that all currently competitive methods to create surrogate models for the prediction of materials properties are not able to capture arbitrary databases evenly yet.
While a fraction of this might be attributed to varying complexity of the databases, the utter failure to capture a ``good'' bandgap-model in the conceptually very simple, large database of cubic perovskites \cite{Castelli2012_cubic} hints that these methods in their commonly used form are not fit to replace DFT to model ``discontinuous'' relationships, where one just replaces a single atom with another compound-unique-species (a finding evident already in the authors previous work \cite{Stanley2019}).
However, for varying ``alloys'' and superstructures in a more or less continous way, such as it happens in the other databases, as well as in \cite{Sutton2019}, the outlined methods seem to be able to perform quite well; a MAE of around $100 \,\unit{meV}$ is great, comparing the inherent inaccuracies of experiments and DFT (GGA vs. hybrids) \cite{Borlido2019}.

Additionally, for all studied descriptors, this study could not establish a strong, order-of-magnitude variation in per-dataset model performance for varying parameters within the boundaries of previously published work, hinting that for all practical applications, a finegrained hyperparameter search \cite{Langer2020} might be inefficient.
Across all datasets, no method consistently reached best performance, though SOAP is leading for several datasets.
Setting aside different modeling techniques for the raw data, the available results for the bandgap models also indicate that choosing a method, much less choosing appropriate parameters for it, has much less influence than choosing a dataset.
Thus these findings question the significance of performing studies on isolated, proprietary datasets aiming for ever better numerical results without establishing baseline performance metrics and a comparison framework \cite{Saidi2020,Park2020,Marchenko2020}.

From a technical perspective, the fact that fingerprinting functions creating input vectors of a length several times the samplesize work so well is a bit puzzling.
Normally one would expect a strong overfitting to the test set, as the models have more free parameters than fitted samples.
While that is exactly the reason for using a regularizing ML method, such as Kernel-Ridge-Regression (KRR), the high sparsity of both the SOAP and MBTR-fingerprint for highly diverse databases could as well mean that the model that way only learns from a fraction of the supplied input data as shown in \cite{Vidovic2015}.
The results of this study, which show SOAP with simple variance-based filtering of input features leading the field, underline this problem.
This should warrant further investigation, as it also means that the given model will never be able to achieve full DFT accuracy just learning the substructure of e.g. O, F and N-atoms, which incidentally are the shared building blocks of the cubic perovskite set, where MBTR excels but has a accuracy which is in a range comparable with compounds swapping the A and B ions \cite{Tao2019}.

It should also be noted that original authors open-sourcing their data or even publishing ML-models should include a recommended training/test-split so that results between methods can be compared across different publications \cite{Wang2020}.
This is especially important, as the usage of neural networks in innovative ways slowly reaches the materials science field. 
With graph neural networks (GNN) already beating the established fingerprinting procedures easily on the GDB-17-derived databases \cite{Gilmer2017, Rahaman2020} so the field can expect further inquiries, for which good and standardized data is a necessary requirement.
The applicability to the solid-state field currently is mostly hampered by small databases of relevant properties as the current, OQMD-derived databases only include formation energies in a consistent, high-quality format and are mostly metallic \cite{Xie2018, Park2019}.
Such developments might also be necessary to escape the fact that actual model performance is more tied to the data than to the model, which looks eerily similar to the state of natural-language-processing 20 years ago \cite{Banko2001}.

The availability of such large-scale databases could also faciliate a more detailed examination of dimensionality reduction and its workings.
While this study shows that the PDDF-fingerprint seems to incorporate information on a low-dimensional manifold for the given datasets and this information in fact allows to construct models of equivalent quality, it is not clear, whether this approach can be further improved and yield extended insights.
t-SNE-analysis hints that the encoding preserves ``chemical information'' while significantly reducing the feature size.
Thus it eases the systematical optimization of the resulting surrogate model in search of new compounds but it is unclear, whether it is thus possible for the model to actually relate to properties of physically realizable compounds.

\section*{Tools}
All calculations were done in a \software{Python}-environment using the \software{numpy}, \software{pandas} and \software{ase} packages for basic data manipulation and structure file handling; plotting was done with \software{matplotlib}.
Machine learning procedures were used/implemented with \software{sklearn} and \software{tensorflow}, while the fingerprints were generated with \software{Dscribe}\cite{Himanen2020} and our own implementation of the \software{PDDF}.
Code to reproduce all numerical experiments is available upon request from the authors.

\section*{Supporting Information}
The supporting information includes a tabular overview of all fingerprint parameters for numerical experiments.
Also, MAEs, R²-scores and RMSEs are included for the subpar-performing fingerprints on the bandgap prediction and for all energy prediction models.

Additional experiments on the data from \cite{Pandey2018} are included to show that for the PDDF, published performance of the fine-grained approach can not be reached with single-property PDDFs alone.

\section*{Acknowledgments}
The authors thank Waldemar Kaiser for a diligent reading of the late draft.

\printbibliography

\end{document}